# Plasmon induced resonant effects on the optical properties of Ag-decorated ZnSe nanowires.

Aswathi K. Sivan[1], Lorenzo Di Mario[2], Daniele Catone[2], Patrick O'Keeffe[3], Stefano Turchini[2], Silvia Rubini[4], and Faustino Martelli[1]

[1] Istituto per la Microelettronica e i Microsistemi (IMM), Consiglio Nazionale delle Ricerche, Via del Fosso del Cavaliere, 100, 00133 Rome, Italy

[2] Istituto di Struttura della Materia-CNR (ISM-CNR), Division of Ultrafast Processes in Materials (FLASHit), Via del Fosso del Cavaliere, 100, 00133 Rome, Italy

[3] Istituto di Struttura della Materia-CNR (ISM-CNR), Division of Ultrafast Processes in Materials (FLASHit), 00016 Monterotondo Scalo, Italy

[4] Istituto Officina dei Materiali (IOM), Consiglio Nazionale delle Ricerche, 34149 Basovizza, Trieste, Italy

e-mail: faustino.martelli@cnr.it

**Abstract** In this work we present how the optical properties of ZnSe nanowires are modified by the presence of Ag nanoparticles on the sidewalls of the ZnSe nanowires. In particular we show that the low-temperature luminescence of the ZnSe nanowires changes its shape, enhancing the phonon-replicas of impurity related recombination and affecting rise and decay times of the transient absorption bleaching at room temperatures, with an increase of the former and a decrease of the latter. In contrast, the deposition of Au NPs on the ZnSe nanowires does not change the optical properties of the sample. We suggest that the mechanism underlying these experimental observations is energy transfer via the Förster interaction based on the fact that the localized surface plasmon resonance (LSPR) of Ag nanoparticles spectrally overlaps with absorption and emission of ZnSe while the Au LSPR does not.

## 1. Introduction.

When photons with a well-defined energy impinge on metallic nanoparticles (NPs), they induce a collective oscillation of the conduction electrons of the NPs giving rise to a localized surface plasmon resonance (LSPR) producing strong resonant fields in its proximity. Plasmons find applications in several fields, from molecular sensing [1] to biomedicine [2] and optoelectronics [3]. Because of the associated electric field, plasmons induced in NPs placed on a semiconductor surfaces may strongly interact with the adjacent material, in particular modifying its optical properties, by means of the Förster interaction (Förster resonant energy transfer, FRET), a non-radiative dipole interaction, [4,5] or by hot carriers transfer from the metallic NPs to the semiconductor [3,6-7]. Modifications of the semiconductor optical properties have been observed in several semiconducting materials when coupled to metal NPs with relevant differences among different material systems [8]. The most common result is the intensity enhancement of the photoluminescence (PL) from the semiconductor that has been reported in quantum wells [9], Si-based light-emitting diodes [10], metal dichalcogenides flakes [11], quantum dots [12-14], and nanowires [5,7, 15-17]. Depending on sample preparation either PL enhancement or quenching have also been observed in a similar system (e.g. Ag-ZnO [18]).

Nanowires (NWs) are characterized by large surface-to-volume ratios thus representing a class of materials whose electronic states can strongly interact with the electric field generated by LSPRs. Metal NPs placed along the semiconductor NWs in high density arrays, form a 3D ensemble in a microporous structure that can have a strong interaction with the environment, thus enhancing the sensing properties of the plasmonic NPs [19]. Moreover, when NPs are obtained by the dewetting of films deposited on the sidewalls of the NWs, the quasi one-dimensional shape of the NW induces NP shapes that differ from the sphere thus modifying their plasmonic properties [20].

In this work, we report the changes induced by Ag NPs on the stationary and transient optical properties of ZnSe NWs. The quasi stationary and dynamical optical properties of the pristine ZnSe NWs have been described in previous publications of our group [21-22] and represent the starting point of the present work. Ag NPs were grown on the sidewalls of the ZnSe NWs by thermal dewetting. The average size of the NPs lies between 20 nm and 50 nm. The LSPR of Ag NPs with those sizes is in the wavelength range of 400-450 nm (3.1-2.7 eV) [23], the range in which the ZnSe band gap (2.7 eV at room temperature [24]) lies. Hence there is a spectral overlap between the LSPR and the absorption band of the ZnSe NWs. In particular, we have studied how the PL and the fast transient absorption in ZnSe NWs are modified by the presence of the Ag NPs. To compare the results of the ZnSe NWs/Ag NPs system with those obtained with a material combination where the LSPR of the NPs is not resonant with the ZnSe band gap, we have also investigated the ZnSe NWs /Au NPs system, with the Au NPs having the LSPR at about 2.34 eV (530 nm) [20], below the ZnSe band-gap. For a similar system (ZnSe:Sb nanoribbons/Si nano-heterojunction) Wang and co-workers [25] have reported the enhancement of the optical absorption when Ag nanospheres are deposited on the ZnSe nanoribbons with the resulting improvement of a light-absorption based optoelectronic device based on these materials.

## 2. Experimental

The ZnSe NWs have been grown by molecular beam epitaxy (MBE) on GaAs, quartz and sapphire at 300 °C. The choice of a low temperature growth was dictated by the resulting superior optical properties of the [21]as high growth temperatures degrade the optical properties of ZnSe with time because of the presence of point defects [26]**.** The samples consist of a very high density of randomly oriented ZnSe NWs. They are tapered, with a mean diameter at the tip of 10 nm and base diameter 10 times larger, with an average length of about 1 μm. No shell protects the ZnSe NWs. Full details on the growth conditions, morphological and structural characterization as well as low-temperature luminescence can be found in Zannier *et al.* [21]**,** while a more complete description of the optical properties, including transient absorption, has been published by L. Tian et al. [22].

Thin films of silver (and gold for comparison) with nominal thickness below 5 nm have been evaporated onto the ZnSe NWs to form NPs after annealing at 280 °C for 10 minutes. For the comparison between pristine NWs and metal-decorated NWs we have used different pieces of the same as-grown sample. Transient absorption measurements have been performed on as-grown samples on quartz or sapphire because the substrates are transparent in the spectral region of interest. PL has been carried out on NWs grown on all three substrates mentioned above. Several samples have been used to verify the reproducibility of the results after metal deposition. We have not observed major dependences on the substrate that will be explicitly mentioned only when minor differences will be pointed out.

Figure 1 shows a scanning electron microscopy (SEM) image of (a) pristine NWs and (b) decorated NWs showing that the Ag NPs have an average diameter of 20-50 nm.

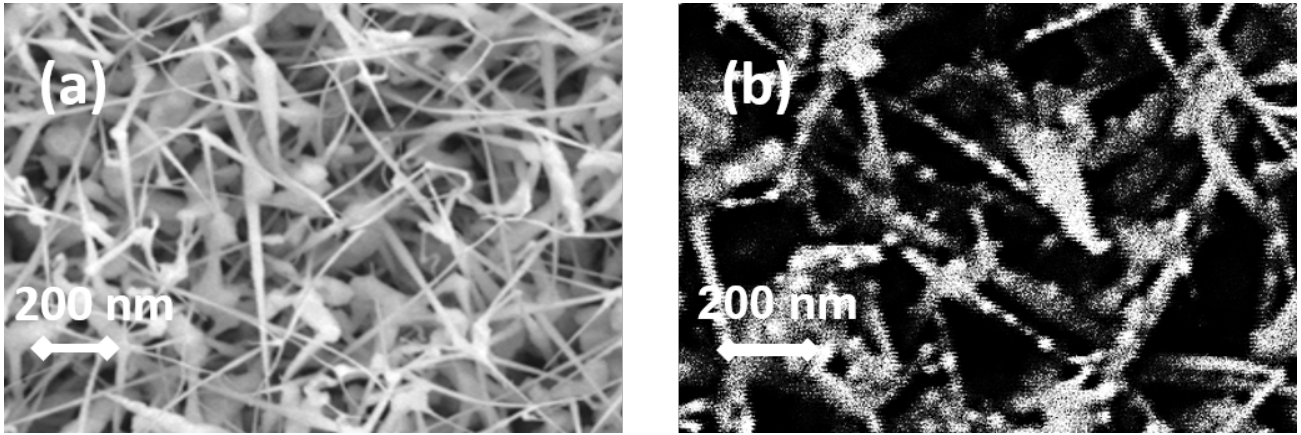

Figure 1. SEM image of typical (a) pristine, and (b) Ag decorated ZnSe NWs.

In figure 2(a) we report the absorbance of Ag NPs fabricated on the sidewalls of silica NWs that show similar size and shape distribution to that of the NPs fabricated on ZnSe NWs. In this type of samples, the absorbance is only due to the Ag NPs. [19,27] In figure 2(b) we show the room temperature PL of a typical ZnSe NW sample used for this work. The maximum absorbance for the Ag NPs is very close in energy to the near band edge (NBE) emission of ZnSe. The doublet structure in the Ag NP absorbance is due to the presence of both dipole and quadrupole contributions.[27]

The PL and time-resolved PL have been performed using a narrow-line laser diode and the second harmonic of a Ti:sapphire laser with a pulse length of 20 fs at 80 MHz repetition rate, respectively. In both cases the excitation wavelength was 405 nm. The PL signal was analyzed using a 0.35 m long monochromator and a charge-coupled device (CCD) or a Si-APD [full width at half maximum (FWHM)=150 ps] for quasi-stationary or time- resolved analysis, respectively. The fast transient absorption spectra (TA) were measured by pump-probe experiments. As a pump, we used the output of an optical parametric amplifier at 410 nm (3.02 eV), with a pulse length of about 40 fs at a 1 kHz repetition rate as input, and as a probe we used the super-continuum white light generated in a commercial femtosecond transient absorption spectrometer of IB Photonics (FemtoFrame II). The instrument response function was measured to be approximately 50 fs. The pump-probe response was studied in the 420–750 nm wavelength range. The measured quantity is the difference in absorbance (ΔA) between the probe transmitted through the excited sample and that transmitted through the unperturbed sample as a function of the delay time between pump and probe.

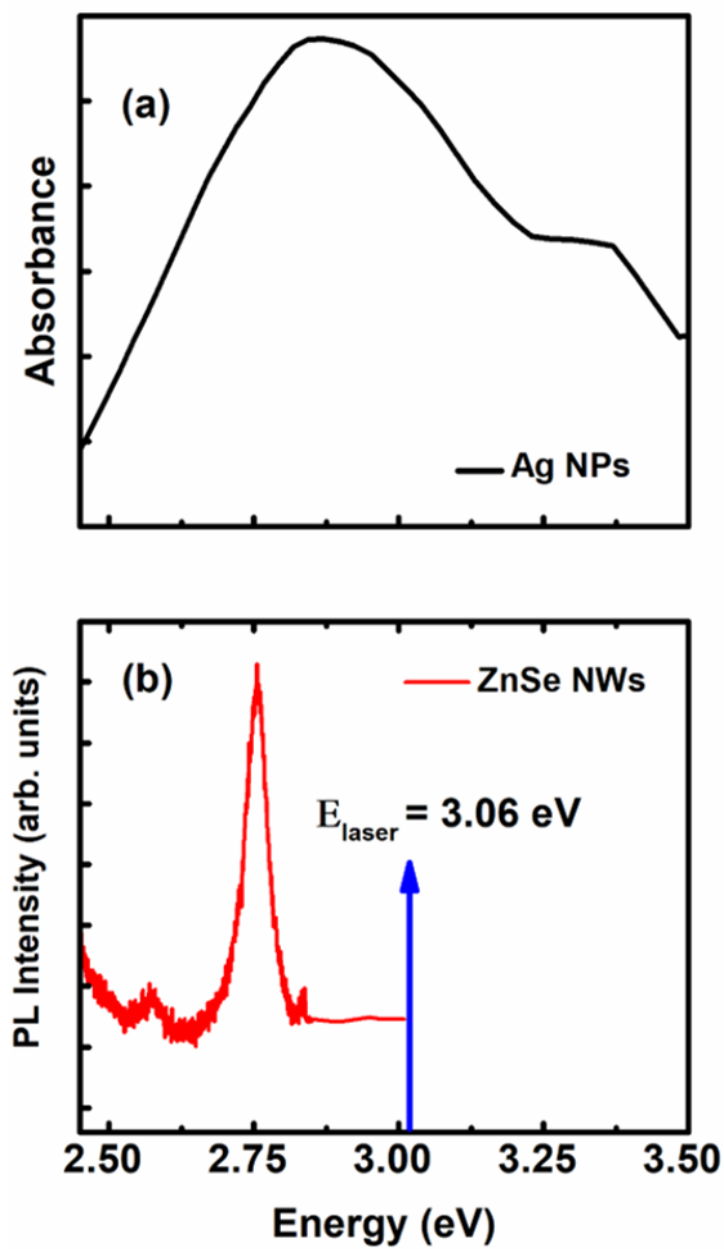

Figure 2. (a) Absorbance of Ag NPs fabricated on the sidewalls of silica NWs, with similar size and shape distribution as the NPs fabricated on ZnSe NWs. (b) Room temperature PL of pristine ZnSe NWs. The blue arrow indicates the pump energy used in the PL measurements (3.06 eV, very close to that used for FTAS namely 3.02 eV).

## 3. Results

### 3.1 Absorption

In figure 3 we show the absorbance of a typical pristine sample (red curve) and its Ag-decorated companion obtained from the same growth (black curve). The absorbance of the Ag-decorated samples is higher than that of the pristine samples and with the line shape of the absorbance that differs from that of the pristine samples especially in the 2.2 eV – 2.6 eV region. The higher absorbance in the decorated samples is due to the added light absorption by the Ag NPs and to an increased light scattering produced by the same NPs.

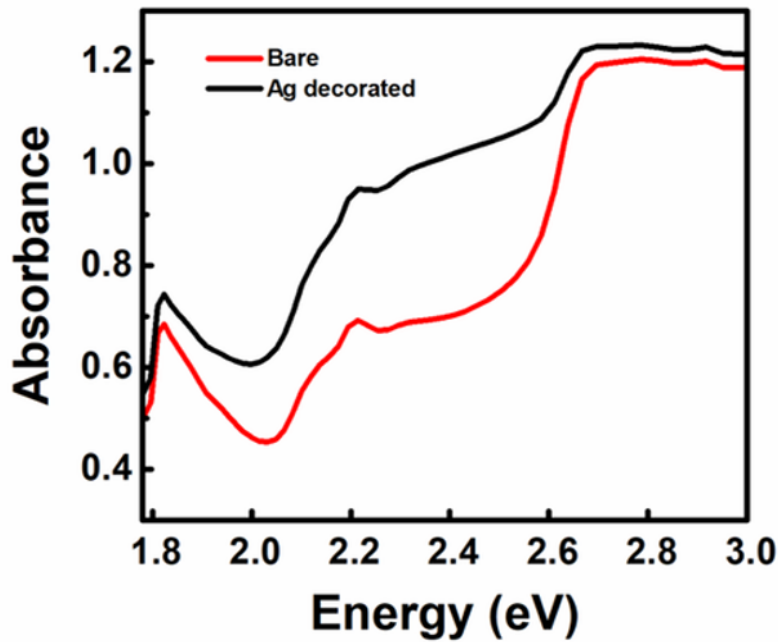

Figure 3. Absorbance of bare (red curve), and Ag-decorated ZnSe NWs (black curve).

In figure 4 we report the 2D map in false colors of the TA spectra in pristine (a) and Ag-decorated samples (b) collected as a function of the pump-probe delay and with a pump fluence of 260 µJ/cm$^2$. In both samples, the bleaching of the absorption of the ZnSe NWs (due to the presence of e-h pairs photoexcited by the pump) is observed at the NBE energy of about 2.7 eV and along the defect band [22]. The two maps appear similar with some differences that we will describe in the following. First, however, we wish to point out the lack of any plasmonic signal in the spectra of the decorated sample.

Ag-NPs of similar size, formed with the same procedure on silica NWs instead show a clear plasmonic signal in the TA spectra [26], in which the contribution from dipole and quadrupole are clearly distinguished.

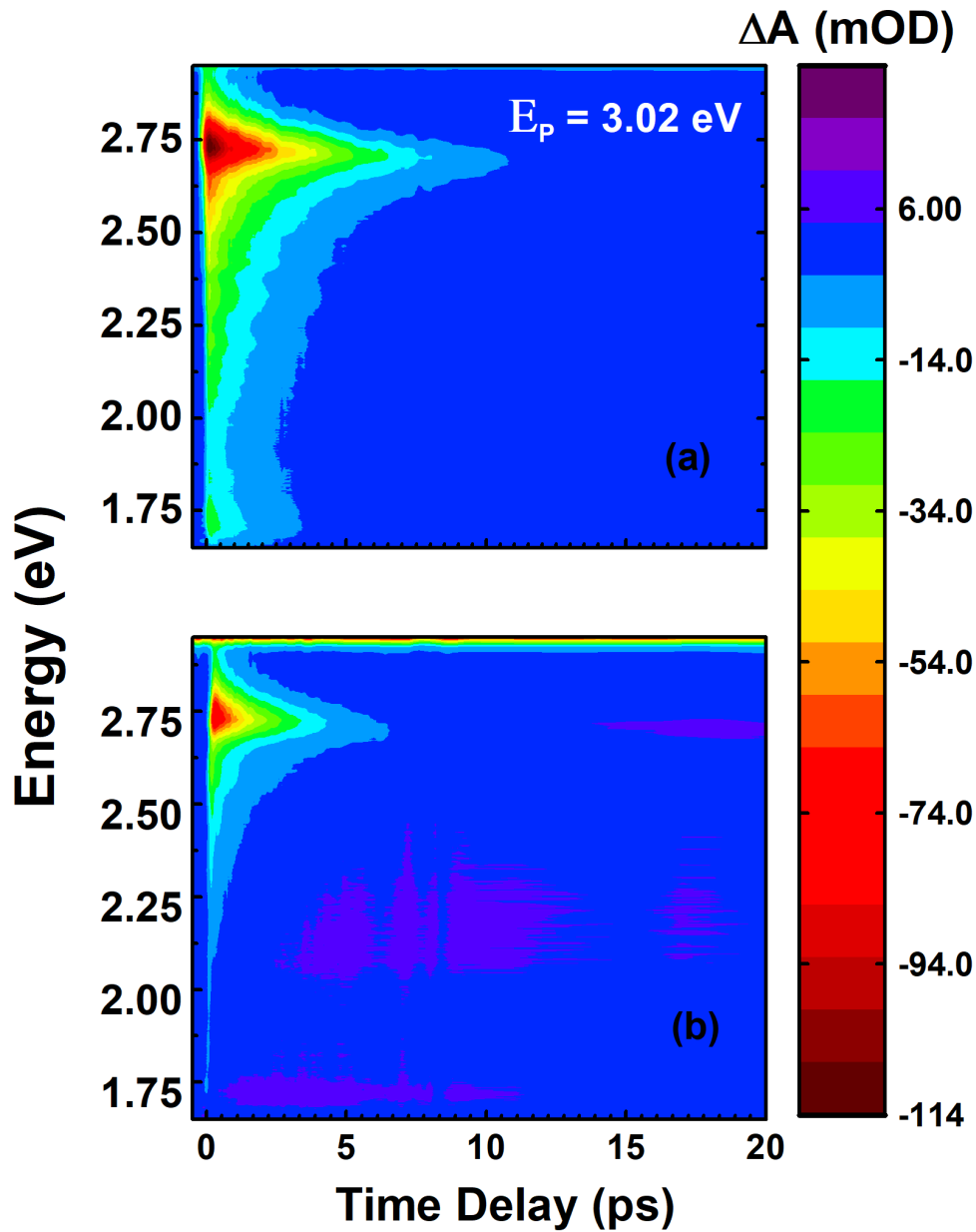

Figure 4. Two-dimensional color map of the absorbance difference, ΔA as a function of the delay time between pump and probe (x axis) and probe energy (y axis) of (a) pristine, and (b) Ag decorated ZnSe NWs.

The intensity of the TA signal of the NBE region [22] in the Ag-decorated sample is reduced by about 20% with respect to that of the pristine sample, a reduction that we attribute to the shadowing caused by the Ag NPs. The main spectral change observed in decorated samples is the reduction of the relative intensity of the defect-related signal at energies below 2.7 eV. This feature can be better appreciated in figure 5, where we report the normalized TA spectra for both type of samples at the same delay time of 1 ps. The TA signal from the defect bands is weaker in the presence of Ag NPs also causing an apparent narrowing of the low-energy side of the NBE bleaching signal.

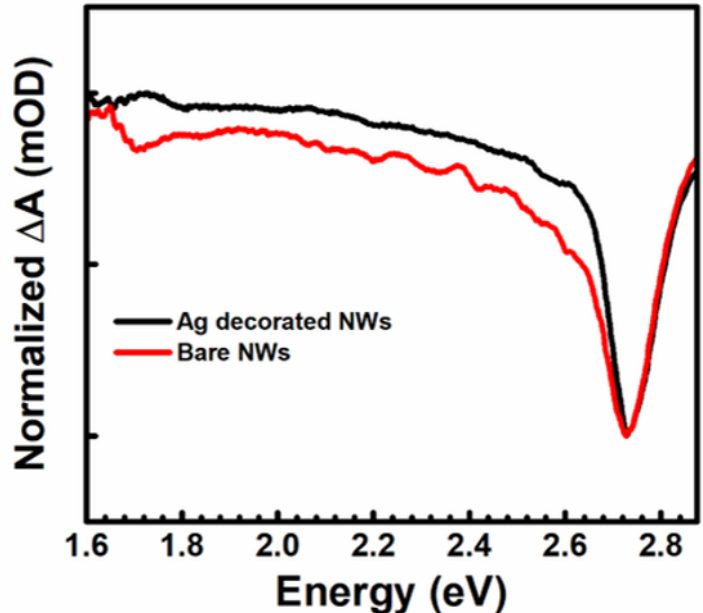

Figure 5. Transient absorption of bare (red curve) and Ag decorated (black curve) ZnSe NWs for a time delay of 1 ps between pump and probe.

A further difference between pristine and Ag-decorated signals can be observed in the dynamics of the NBE signal. In figure 6 we report the temporal behavior of ΔA in both type of samples. In figures. 6(a) and 6(b) we report the risetime of ΔA at the NBE energy in the pristine and Ag-decorated sample, respectively. It increases from $\tau_r=165\pm52$ fs in the pristine sample (within the error bar, this value is the same measured in similar samples in Tian et al.[22]) to $\tau_r=250\pm95$ fs in the Ag-decorated sample. On the other hand, a reduction is observed in the decay lifetime of the TA signal at the NBE energy in the Ag-ZnSe NWs with respect to the bare ZnSe NWs. In the former case we find $\tau_d=1.85\pm0.05$ ps as compared with $\tau_d=3.57\pm0.08$ ps in the latter one. Finally, figure 6(d) shows the values obtained for the decay time as a function of the probe energy: for the pristine samples, a maximum is observed at energies close to the band-gap value. In general, the broad band at low energy exhibits faster dynamics than the band-gap region, as seen in the transient maps in Figure 4. This observation is in agreement with the previous work reported in Tian et al. [22]. This behavior is also observed in Ag decorated NWs, but in this case the decay times are lower than their corresponding values in bare NWs in the whole energy range explored in the present work.

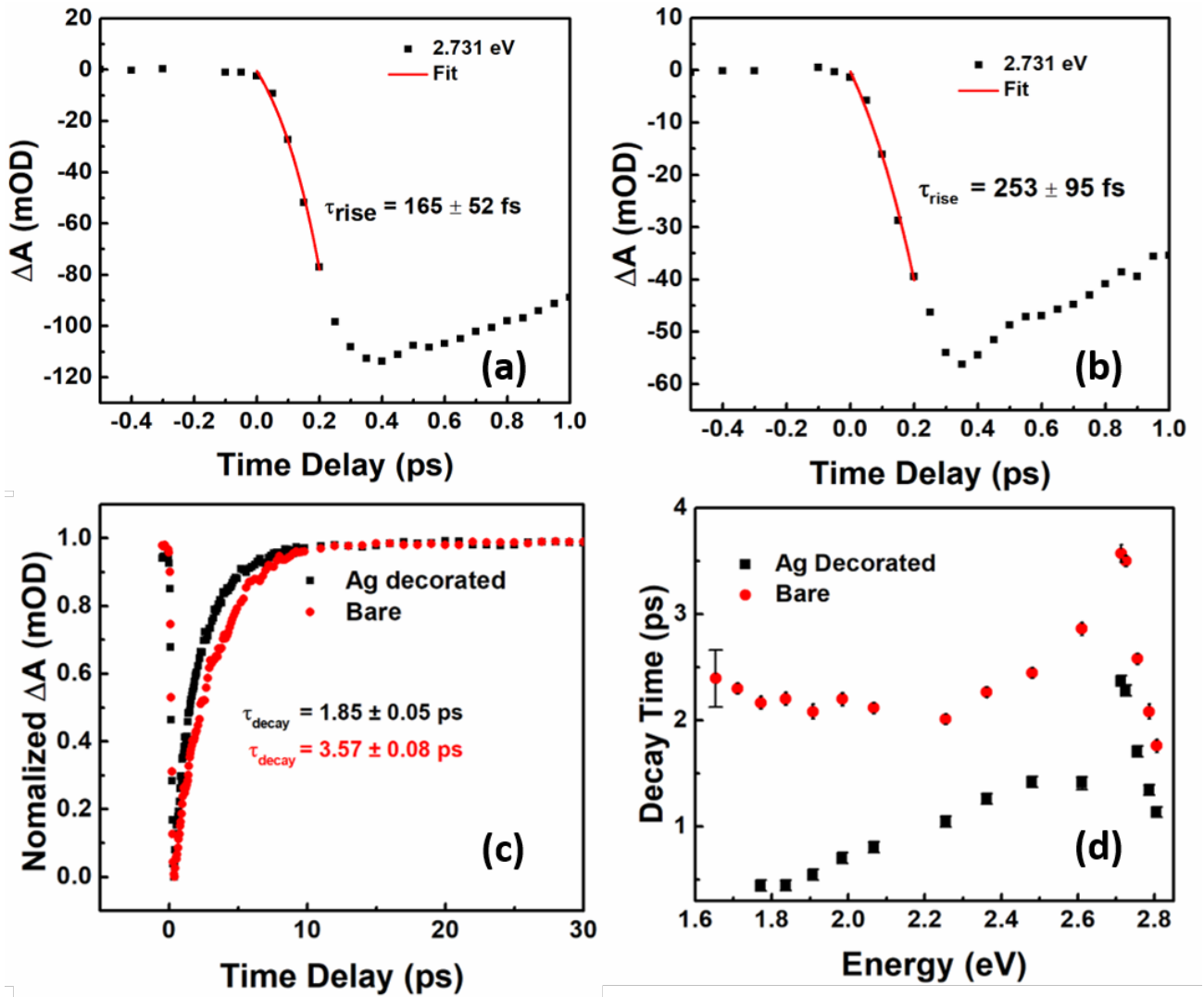

Figure 6. Temporal cut at the bandgap energy (2.73 eV) for (a) pristine and (b) Ag decorated ZnSe NWs, together with the fit trace (red curve) pointing out the rise time of the bleaching signal. (c) Temporal cut at the bandgap energy (2.73 eV) of bare (red curve) and Ag-decorated (black curve) ZnSe NWs. (d) Spectral dependence of the decay time of ΔA of bare (red points) and Ag-decorated (black points) ZnSe NWs.

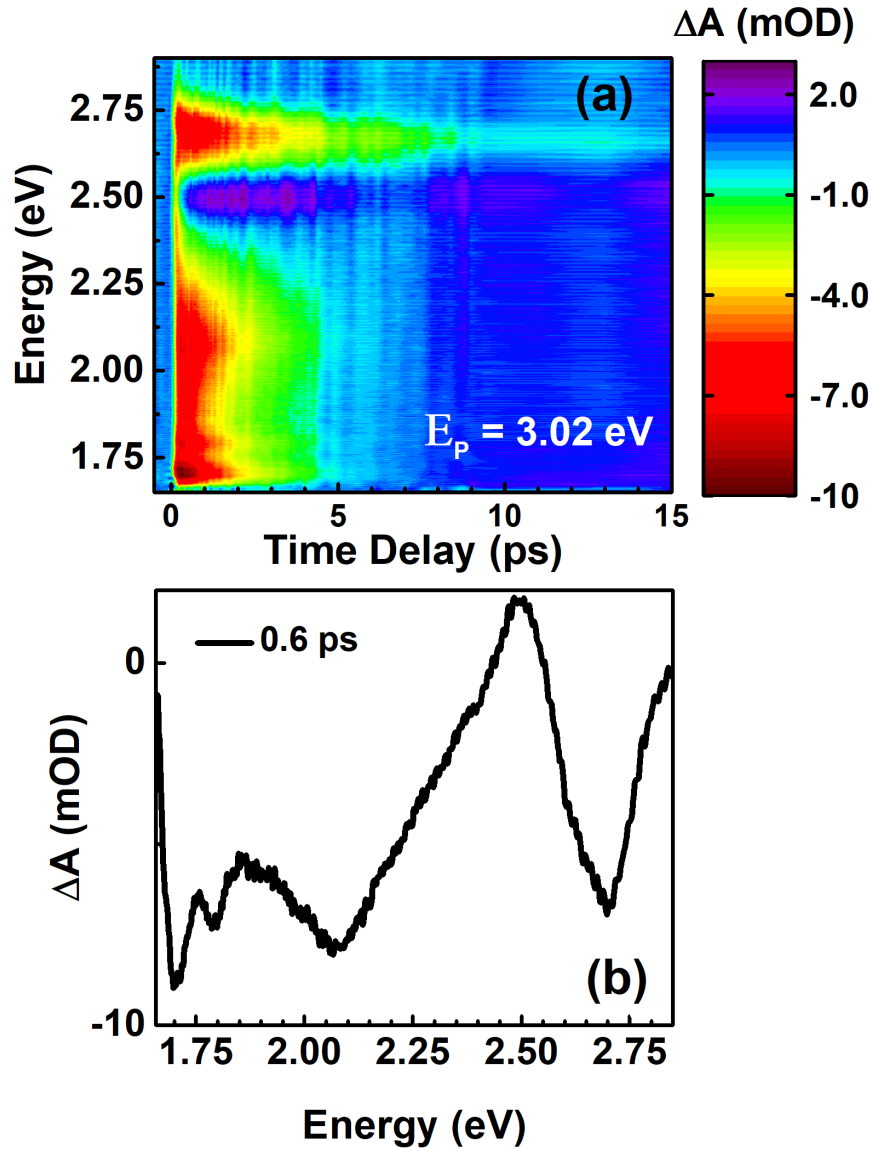

Figure 7. (a) Two-dimensional color map of the absorbance difference, ΔA as a function of the delay time between pump and probe (x axis) and probe energy (y axis) of Au decorated ZnSe NWs. (b) Transient absorption of Au decorated ZnSe NWs at a time delay of 0.6 ps.

As mentioned in the introduction, to point out the importance of having LSPRs with energy close to that of the intrinsic electronic states of the semiconductor in order to have the observed differences, we have also studied the optical properties of ZnSe NWs decorated with Au NPs. In this case the LSPR is not resonant with the band gap of ZnSe NWs [20]. Figure 7 shows the 2D transient map of the ΔA intensity vs time delay and probe energy for Au-decorated ZnSe NWs excited with a pump of 3.02 eV at an excitation intensity of 260 µJ/cm². In these Au-decorated NWs, beyond the bleaching signal at the NBE energy, we can clearly see contributions of the Au LSPR to the TA signal at energies around 2.3 eV. This feature is favored by a lower absorption of the ZnSe NWs in the Au LSPR region than at the energy of the Ag LSPR. As a result, one of the positive wings of the plasmonic signal [20] is clearly visible at 2.5 eV. At lower energy the TA signals of the Au LSPR and that of the ZnSe defects mix with each other (Figure7(b)). The most important thing to notice, however, is that the TA dynamics of the NBE signal in Au-decorated samples does not differ from that observed in the pristine samples. The decay time in case of the Au-decorated NWs is $\tau_d$=3.69 ± 0.21 ps and a rise time of $\tau_r$=150 ± 90 fs. If we use a pump energy resonant with the Au plasmon (2.34 eV, 530 nm), which is smaller than the ZnSe band gap, we observe a very weak bleaching of the NBE absorption in all samples, pristine or metal decorated, because of the partial absorption of the pump by defect states in ZnSe [22] that excites electrons in the conduction band. The bleaching signal appears very similar in all samples.

## 3.2 Photoluminescence

The PL spectra of ZnSe at 10 K typically show three emission bands corresponding to the NBE region, to shallow centers and to the Y-line at low energies.[21,28] ZnSe being a polar crystal, because of Fröhlich interaction, phonon replicas often appear in the PL spectra from phonon-assisted emission of free and bound excitons as well as donor-acceptor pair (DAP) [28].

Figure 8 reports the PL spectra at 10 K of our ZnSe NWs with and without the decoration of Ag and Au NPs. The intensity of the NBE PL in the decorated samples is reduced with respect to that of the pristine samples because of shadowing of the pump and of the PL by the metal NPs. The reduction is of about the same magnitude in Ag- or Au-decorated samples. In Ag-decorated NWs, the relative intensity of the Y-line decreases with respect to the NBE emission, while the emission of the energy region right below the NBE increases.

In particular, the spectra of the pristine sample (red lines in parts (a), (b), and (c) of Figure 8) are characterized by a broad NBE emission peaked at about 2.8 eV followed by a broader emission at low energies (the Y-line, see also ref. **21**). Depending on the substrate used, GaAs, quartz or sapphire, some differences can be observed from sample to sample in terms of NBE width and its relative intensity with respect to the broader low-energy emission. In particular, for the sample grown on sapphire, the broad impurity-related emission is much stronger than the NBE emission, probably because of contamination from the substrate. Independently of the relative intensity of defect and NBE signals, in all samples the deposition of Ag NPs dramatically changes the line shape of the PL spectra (black lines in parts (a), (b), and (c) of Figure 8). The broad emission around 2.76 eV splits into several narrower peaks. We distinguish the NBE at 2.81 eV, while in the energy range between 2.6 and 2.76 eV, 4 to 5 (depending on sample) equidistant peaks are observed with energy separation equal to the LO phonon energy of ZnSe (31.5 meV).

We attribute the equally separated peaks observed in the Ag-decorated samples to no-phonon (at the highest energy) and phonon replicas of the impurity-related recombination (bound exciton and/or donor-acceptor pairs)[28-31]. The formation of Ag NPs on the ZnSe NW sidewalls gives rise to a strong enhancement and a narrowing of the emissions relative to the impurity related recombination, also allowing the observation of distinct phonon replicas.

We point out that Au NPs deposited on the ZnSe sidewalls do not cause any significant change to the PL line shape (green lines in Figs. 8 (a) and (b)), giving only rise to a decrease of the NBE PL intensity due to bare shadowing of illumination and light emission because of the presence of the light-absorbing Au NPs. The defect-related emission in the energy range 2.6-2.75 eV is slightly reduced relative to the NBE emission, while the Y-line is slightly enhanced.

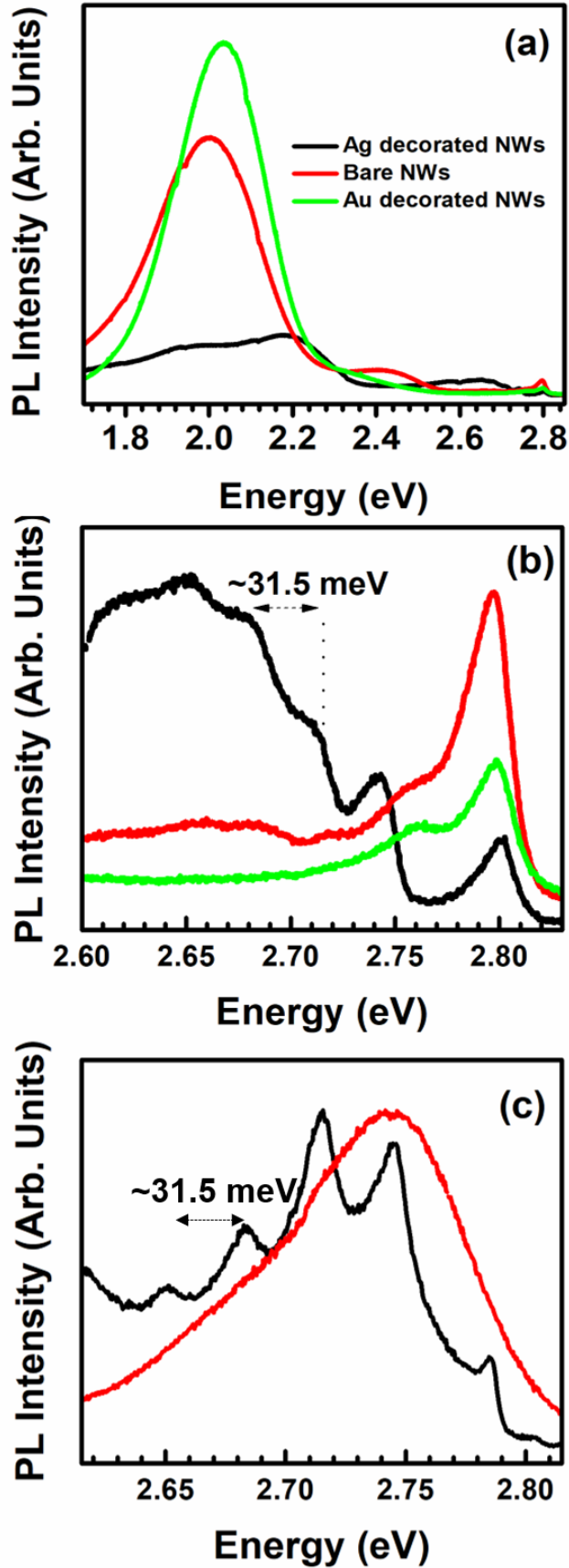

Figure 8. (a) PL spectra of bare (red) and Ag decorated (black) and Au-decorated (green) ZnSe NWs grown on quartz. (b) Detail of the high-energy part of the PL spectrum shown in (a); (c) Normalized PL spectra of bare(red) and decorated(black) ZnSe NWs grown on sapphire, showing the DAP region. All spectra recorded at a lattice temperature of 10 K.

Time resolved measurements of the NBE luminescence in all samples (not shown) do not indicate any significant change in the recombination lifetime (about 0.5 ns) between decorated and pristine samples, in agreement with the invariance of the PL intensity.

4. Discussion

Electronic interaction between a semiconductor and a plasmonic nanostructure can mainly occur through two different pathways. These pathways depend upon the spatial proximity and the relative energy position of the LSPR of the plasmonic nanoparticles and the semiconductor electronic states. As mentioned in the introduction, these two mechanisms, namely hot electron injection and FRET, benefit from the proximity between the involved materials, but only for FRET a further necessary condition is the spectral overlap between the LSPR and the absorption band of the semiconductor[4]. In our system of Ag-decorated ZnSe NWs, there are spectral overlap and physical contact between the metal and semiconductor NW. This could enable both the FRET and hot electron transfer to be possible. In case of Au decorated ZnSe NWs, there is a physical contact between the metal and semiconductor, but no spectral overlap between Au LSPR and ZnSe band gap absorption, hence only hot charge carrier transfer could be observed in our experiment.

The decay of the NP LSPR produces hot electrons in both Ag and Au NPs, whose energy distribution will reflect the density of states in the two metals. Because of the band alignment between the metals and ZnSe [32-33], part of the hot electrons will have energy above the ZnSe conduction band. It is worth pointing out that the electronic density of states at the Fermi level is small for both metals. In both cases, indeed, the Fermi level of the metals lies well above the center of the relative electronic *d* band, 2.58 eV for Au [34] and 3.93 eV for Ag [35]. Given the presence of a Schottky barrier between the metals and ZnSe [33], in both metals only electrons close to the Fermi level will have sufficient excess energy to overcome it. Then, for the electronic conditions we have very close similarities between the Ag/ZnSe NW and Au/ZnSe NW interfaces. Therefore, we argue that if the changes in the optical properties of the Ag-decorated NWs were due to hot electron transfer from the metal NP to the semiconductor we should observe similar modifications to our TA and PL spectra with both metals, which is not the case. Hence, we suggest that the mechanism underlying the modifications observed in the TA and PL spectra after the formation of the Ag NPs on the NW sidewalls is the Förster interaction.

The increase in the rise time of the bandgap bleaching in the Ag-decorated ZnSe NWs may indicate the presence of FRET mechanism where the Ag NPs act as donors and the ZnSe NWs are the acceptors [4,36]. The increase of the rise time could be explained as due to the excitation of e-h pairs in ZnSe in a later time than for direct photoexcitation, as the result of the energy transfer from the NPs toward the NWs. The decrease of the decay time of the absorption bleaching is observed for the whole energy range from NBE down to the defect band. The absorption spectrum of the Ag NPs is very broad due to the presence of both dipole and quadrupole excitation of the plasmon and of the shape distribution of the NPs [20, 24, 37-38]. At low energies free-carrier absorption also contribute to their absorbance spectrum. The broad width may allow the resonant interaction with the defect band with effects similar to that observed for the NBE. The more efficient coupling of the carriers to phonons suggested by the low-temperature PL of the Ag decorated NWs (see below) may also favor carriers to lose their initial energy towards electronic states at lower energy. An increased carrier-phonon scattering rate would explain the faster decay time in the TA signal of the Ag-decorated samples. Furthermore, the experimental observation of the plasmonic bleaching in the Au/ZnSe NW transient absorbance spectrum but not in the Ag/ZnSe NW case could indicate that the energy transfer in the latter system is very efficient, i.e. the fact that the Ag-NPs are not significantly heated suggests that the energy absorbed by the plasmon resonance is efficiently transferred to the ZnSe NWs. Of course, we cannot rule out that the Ag-LSPR bleaching could be small with respect to the ZnSe-NBE bleaching, but we notice that the bleaching of the Ag-LSPR in a similar 3D array of Ag NPs on silica NWs has an intensity of about 30 mOD [27], of the same order of the ZnSe-NBE bleaching observed here. Therefore, the absence of a clear contribution of the Ag-LSPR to the TA signal of the Ag/ZnSe NW system points towards a very efficient FRET picture as suggested a few rows above.

Less clear is to understand is how the Förster interaction could act in changing the PL line shape in the Ag-decorated ZnSe NWs and, in particular, the enhancement and narrowing of the impurity-related recombination with its phonon replicas. Nevertheless, the Förster interaction is a possible origin of the observed features, in particular favoring the carrier-phonon coupling. Indeed, we can rule out a simple passivation mechanism due to the presence of the Ag NPs on the NW sidewalls because not the smallest hint of line shape changes of the same type is observed with Au-decorated NWs. Noticing that the NP coverage of the NWs is very similar for the two cases of Ag and Au we can also exclude that the differences are merely due to a different degree of surface passivation. The impurity-related recombination only occurs at low temperatures because of the involvement of shallow donors and acceptors [39] and related effects cannot be observed at room temperature at which we perform our FTAS measurements. This feature does not allow us to make direct comparison between the low-T PL and the RT FTAS. Nevertheless, both measurements point towards a more efficient carrier-phonon interaction rate. On the other hand, Raman measurements performed at RT (not reported) do not show any relevant difference between the spectra of bare and Ag- NWs. A reduction of the overall intensity of the Raman peaks of the decorated NWs is the only observed feature, due to shadowing of ZnSe by the metal NPs.

## 5. Conclusion

We have reported the changes of transient absorption and luminescence of ZnSe nanowires induced by the fabrication of Ag nanoparticles on the nanowire sidewalls. In particular we have observed the change of rising and decay times characteristic of the dynamics of the absorption bleaching in the nanowires and the dramatic changes in the low-temperature luminescence spectra, where a strong enhancement of the phonon mediated impurity-related recombination has been observed. On the other hand, the coverage of the ZnSe NWs with Au nanoparticles do not lead to any significant change in transient absorption and luminescence of the ZnSe. We have attributed the observed feature to the Förster interaction, which is favored in the Ag-ZnSe system with respect to that of the Au-ZnSe system because of the near energy resonance of Ag localized surface plasmon resonance and ZnSe band gap. We hope that our experimental results will stimulate further investigation including theoretical calculations.

**Acknowledgments.** This paper has received funding from the Horizon 2020 program of the European Union for research and innovation, under grant agreement no. 722176 (INDEED). This is the version of the article before peer review or editing, as submitted by an author to Nanotechnology. IOP Publishing Ltd is not responsible for any errors or omissions in this version of the manuscript or any version derived from it. The Version of Record is available online at https://doi.org/10.1088/1361-6528/ab68ba